\documentclass[
aps,
prx,
twocolumn,
superscriptaddress,
floatfix,
longbibliography,
nofootinbib
preprintnumbers
]{revtex4-2}
\usepackage[dvipdfmx]{graphicx}
\usepackage{dcolumn}
\usepackage{bm}
\usepackage{ulem}
\usepackage{amsmath}
\usepackage{amssymb}
\usepackage{txfonts}
\usepackage{hyperref}
\usepackage{color}
\usepackage{xcolor}
\usepackage{listings}
\usepackage{cprotect}

\usepackage{tikz}
\usetikzlibrary{quantikz2}

\graphicspath{{./figures/}}

\newcommand{\Tr}{{\rm Tr}}
\newcommand{\e}{{\rm e}}

\newcommand{\im}{\mathrm{i}}

\newcommand{\mF}{\mathrm{F}}

\hypersetup{
  colorlinks=true,
  linkcolor=[rgb]{0.60,0.00,0.00},
  citecolor=[rgb]{0.00,0.00,0.60},
  urlcolor=[rgb]{0.00,0.00,0.60},
  setpagesize=false
}

\definecolor{codegreen}{rgb}{0,0.6,0}
\definecolor{codegray}{rgb}{0.5,0.5,0.5}
\definecolor{codepurple}{rgb}{0.58,0,0.82}
\definecolor{backcolour}{rgb}{0.95,0.95,0.92}

\lstdefinestyle{mystyle}{
    backgroundcolor=\color{backcolour},   
    commentstyle=\color{codegreen},
    keywordstyle=\color{magenta},
    numberstyle=\tiny\color{codegray},
    stringstyle=\color{codepurple},
    basicstyle=\ttfamily\footnotesize,
    breakatwhitespace=false,         
    breaklines=true,                 
    captionpos=b,                    
    keepspaces=true,                 
    numbers=left,                    
    numbersep=5pt,                  
    showspaces=false,                
    showstringspaces=false,
    showtabs=false,                  
    tabsize=2
}

\begin{document}

\preprint{RIKEN-iTHEMS-Report-25}

\title{
Digital quantum simulation of many-body localization crossover in a disordered kicked Ising model
}

\author{Tomoya Hayata}
\email{hayata@keio.jp}
\affiliation{Departments of Physics, Keio University School of Medicine, 4-1-1 Hiyoshi, Kanagawa 223-8521, Japan}
\affiliation{RIKEN Center for Interdisciplinary Theoretical and Mathematical Sciences (iTHEMS), RIKEN, Wako 351-0198, Japan}
\affiliation{International Center for Elementary Particle Physics and The University of Tokyo, 7-3-1 Hongo, Bunkyo-ku, Tokyo 113-0033, Japan}

\author{Kazuhiro~Seki}
\email{kazuhiro.seki@riken.jp}
\affiliation{Quantum Computational Science Research Team, RIKEN Center for Quantum Computing (RQC), Saitama 351-0198, Japan}

\author{Seiji Yunoki}
\email{yunoki@riken.jp}
\affiliation{Quantum Computational Science Research Team, RIKEN Center for Quantum Computing (RQC), Saitama 351-0198, Japan}
\affiliation{Computational Quantum Matter Research Team, RIKEN Center for Emergent Matter Science (CEMS), Wako, Saitama 351-0198, Japan}
\affiliation{Computational Materials Science Research Team, RIKEN Center for Computational Science (R-CCS), Kobe, Hyogo 650-0047, Japan}
\affiliation{Computational Condensed Matter Physics Laboratory, RIKEN Pioneering Research Institute (PRI), Saitama 351-0198, Japan}

\begin{abstract}
Simulating nonequilibrium dynamics of quantum many-body systems is one of the most promising applications of quantum computers.
However, a faithful digital quantum simulation of the Hamiltonian evolution is very challenging in the present noisy quantum devices. 
Instead, nonequilibrium dynamics under the Floquet evolution realized by the Trotter decomposition of the Hamiltonian evolution with a large Trotter step size is considered to be a suitable problem for simulating in the present or near-term quantum devices.
In this work, we propose simulating the many-body localization crossover as such a nonequilibrium problem in the disordered Floquet many-body systems.
As a demonstration, we simulate the many-body localization crossover in a disordered kicked Ising model on a heavy-hex lattice using $60$ qubits from $156$ qubits available in the IBM Heron r2 superconducting qubit device named ibm\_fez. 
We compute out-of-time-ordered correlators as an indicator of the many-body localization crossover. 
From the late-time behavior of out-of-time-ordered correlators, we locate the quantum chaotic and many-body localized regimes as a function of the disorder strength.
The validity of the results is confirmed by comparing two independent error mitigation methods, that is, the operator renormalization method and zero-noise extrapolation.

\end{abstract}

\date{\today}

\maketitle


\section{Introduction}
\label{sec:introduction}
The phases of quantum many-body systems driven out of equilibrium have been intensively studied in recent years.
Among them, periodically driven systems -- so-called Floquet systems -- have attracted growing attention because they can exhibit exotic phases that are inaccessible in equilibrium systems~\cite{Bruno2013,Watanabe2015} such as time crystals~\cite{PhysRevLett.109.160401,Sacha2015} and time quasi-crystals~\cite{Pizzi2019}.

Thermalization is ubiquitous in quantum many-body systems. Even driven quantum many-body systems typically thermalize~\cite{Alessio2013} as a low-energy, low-entanglement state absorbs energy from the driving force and evolves into a high-energy, highly entangled state that may obey the eigenstate thermalization hypothesis~\cite{Deutsch1991,Srednicki1994,Rigol2008}. 
Under thermalization, the information encoded in a quantum state is spread across the entire system irreversibly, causing coherent oscillations to decay.
Hence, a stable quantum-time crystal demands a mechanism to protect the system from thermalization.

The so-called many-body localization (MBL)~\cite{Znidaric2008,Pal2010} is a robust mechanism that prevents disordered quantum many-body systems from thermalization~\cite{Nandkishore2015,Abanin2019,Sierant_2025}.
The strong disorder completely alters the properties of the energy eigenstates, causing the breaking of the eigenstate thermalization hypothesis. 
It is known that the introduction of disorder in Floquet systems can also lead to the MBL:
the Floquet Hamiltonian that describes the spectral properties of the driven system exhibits the transition from an infinite-temperature thermal phase, which obeys the eigenstate thermalization hypothesis, to an integrable many-body localized phase that does not~\cite{Ponte2015,PLazarides2015,Zhang2016}.

Trotterization is a standard method for simulating the time evolution of quantum many-body systems using digital quantum computers.
It approximates the time evolution operator $e^{-\im t H}$ by a product of exponentials composed of noncommutative subsets of the Hamiltonian $H$.
Importantly, the Trotter dynamics, that is, the time evolution realized by applying the Trotterized time evolution operator repeatedly, can be understood as the Floquet systems in which the time-dependent Hamiltonian switches in the noncommutative subsets of the Hamiltonian by identifying the Trotter time step as the driving period~\cite{Heyl2019}.
This equivalence may enable us to study the time evolution of some Floquet system more easily than that of Hamiltonian systems using digital quantum computers~\cite{Kim2023,Yang2023,Shinjo2024,Seki2025,Hayata2024,Hayata2025}.
In fact, it was argued that quantum simulation of the prethermal state of Floquet systems with a classically intractable system size would be one of the best targets in near-term digital quantum computers~\cite{Yang2023}.

As aforementioned, the MBL is a dynamical phenomenon that prevents the system from thermalization, and the dynamical transition between the thermalizing and ergodicity-breaking localized phases~\cite{Sahay2021}, that is, the MBL transition would be hard to simulate in classical computers except small systems because we need to simulate nonequilibrium dynamics or full eigen spectrum.
Hence, the quantum simulation of the MBL transition in Floquet systems with a classically intractable system size would also be a promising target for near-term digital quantum computers.
Indeed, the MBL  has been studied with superconducting digital quantum computers by examining the thermalization of local observables ~\cite{Mi2021TC} and identifying the local integral of motion operators~\cite{Shtanko2025}, or using the excited-state variational quantum eigensolver~\cite{Liu2024}.
However, the effects of disorder on information scrambling -- a phenomenon that characterizes how ergodicity emerges or fails in isolated quantum systems -- have yet to be experimentally investigated using digital quantum computers.

In this paper, we make the above idea more convincing by demonstrating a digital quantum simulation of the MBL crossover in a periodically driven quantum many-body system of a fixed number of qubits.
To this end, we compute out-of-time-ordered correlators (OTOCs) as an indicator of the MBL crossover~\cite{Huang2016,Chen2016,ChenXiao2016,Fan2017,Smith2019}. 
OTOCs are a measure of the information scrambling: It quantifies the spreading of local information thrown to the quantum state, which is intimately related to thermalization.
Hence, OTOCs change qualitatively in the chaotic and many-body localized (ergodicity-breaking) phases, and they may be used as an indicator of the MBL crossover.
In fact, as will be shown, we find that the late-time behavior of OTOCs can be used as an indicator of the MBL crossover.
Specifically, we compute OTOCs in a disordered kicked Ising model on a heavy-hex lattice using $60$ qubits from $156$ qubits available in the IBM Heron r2 quantum processor named ibm\_fez. 
The kicked Ising model on a heavy-hex lattice can be most naturally implemented on IBM's Heron quantum processor due to its compatibility with the qubit layout and the native two-qubit gate of the device~\cite{Kim2023}.

The present quantum computers are necessarily subject to noise, and we need to mitigate the noise effect to obtain reliable results using them.
Here, we apply two error mitigation methods for cross-check: One is the operator renormalization and the other is zero-noise extrapolation (ZNE).
As detailed below, we may find a natural reference circuit to estimate the decay of the fidelity, that is, the renormalization factor of OTOCs~\cite{Swingle2018,Mi2021,Seki2025}, 
while the latter is the standard method in error mitigation.
We demonstrate that the present digital quantum computers, with the help of error mitigation, have the capability of simulating nonequilibrium dynamics of Floquet many-body systems consisting of over 50 qubits, which is not feasible in the exact diagonalization methods with classical computers.
Our work may pave a new direction in near-term applications of digital quantum computers in the era before the realization of fault-tolerant quantum computers.

\section{Disordered Kicked-Ising model}
\label{sec:model}

\begin{figure}
  \includegraphics[width=.48\textwidth]
  {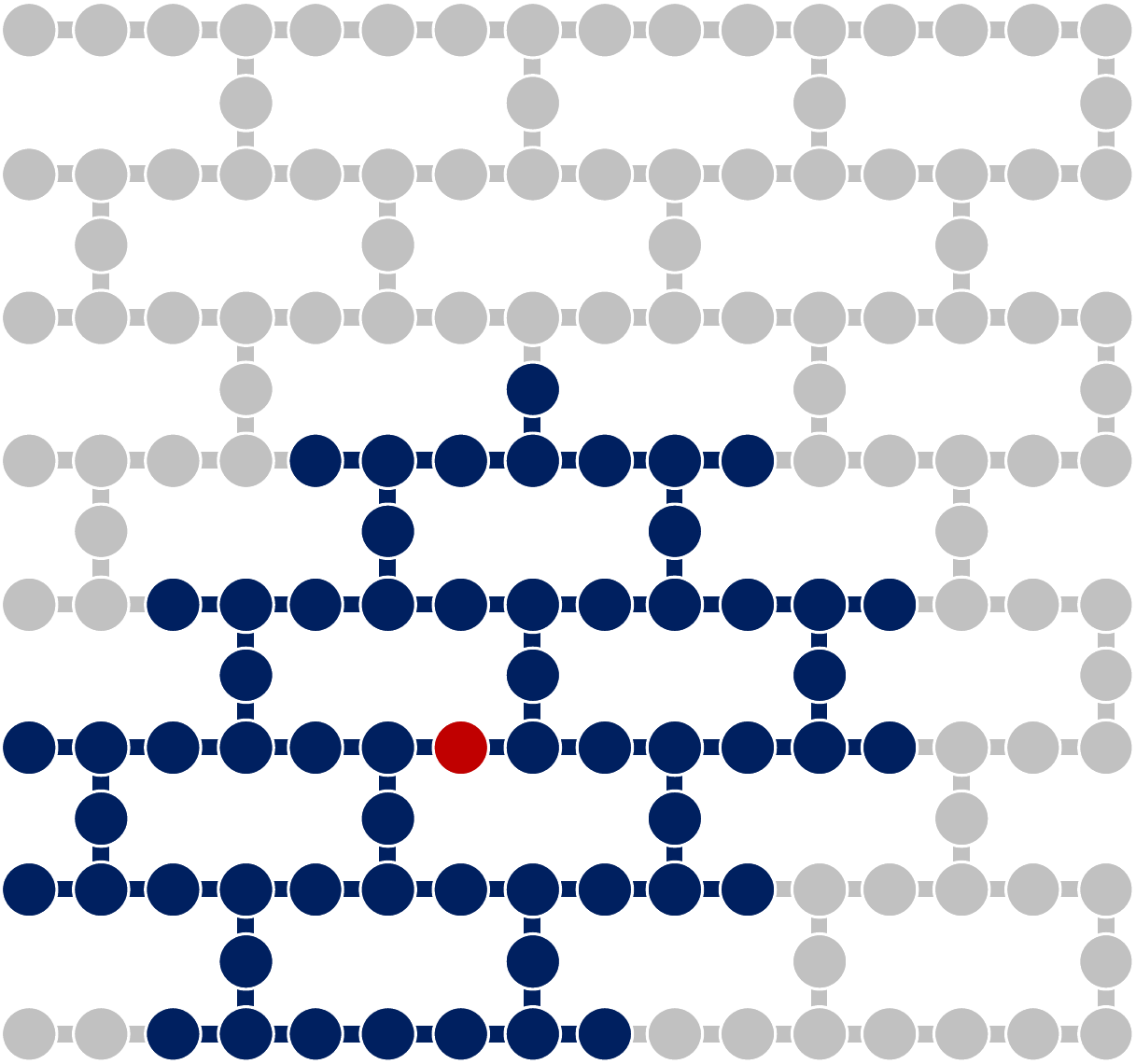}
  \caption{
    \label{fig:qubit_layout}Layout of qubits in ibm\_fez. The vertices colored in dark blue and red denote the qubits used in our experiments. The vertex colored in red denotes the position of the butterfly operator. The edges denote the connectivity of qubits in IBM's Heron r2 devices.
  }
\end{figure}

We study the disordered kicked Ising model on a heavy-hex lattice, which is suitable for simulation in IBM quantum devices. The time evolution of the model is described by the following time-dependent Hamiltonian:
\begin{align}
\label{eq:periodic_H}
  &H(t) =
  \nonumber\\
  &\left\{
    \begin{array}{lll}
      \displaystyle{H_X = \sum_{i=1}^{N}B_X(i) X_i,}
      & \displaystyle{t\in [0, T/2)},
      \\
      \displaystyle{H_Z = J\sum_{(i,j)}Z_iZ_{j} + B_Z\sum_{i=1}^{N} Z_i,}
      & \displaystyle{t\in[T/2,T)},
    \end{array}
  \right.
\end{align}
where $X_i$ ($Z_i$) is the component $x$ ($z$) of the Pauli operator at the
site $i$ and $N$ is the total number of sites. 
The sum of $ZZ$ interaction terms in $H_Z$ is taken over the nearest-neighbor sites $(i,j)$ in the heavy-hex lattice (See Fig.~\ref{fig:qubit_layout}). 
While the coefficients $J$ and $B_z$ in $H_Z$ are fixed, the site-dependent coefficient $B_X(i)$ in $H_X$ is a random field and uniformly distributed in $[B_{X0}-W,B_{X0}+W]$, with $B_{X0}$ and $2W$ being the center and interval, respectively.

The time evolution operator over a single driving period, i.e., the Floquet operator reads 
\begin{equation}
\label{eq:U_F}
    U_\mathrm{F}
    =
    \e^{-\im H_Z T/2} \e^{-\im H_X T/2}.
\end{equation}
The evolved state at time $nT$ ($n=0,1,2\ldots$) is exactly given by $|\psi(nT)\rangle=(U_\mathrm{F})^n|\psi(0)\rangle$ with $|\psi(0)\rangle$ being the initial state.
We fix $JT=\pi/2$, $B_ZT=1.3$, and $B_{X0}T=\pi/2$. 
Then, the model is maximally chaotic in the absence of the disorder, that is, $W=0$~\cite{Bertini2018SFF, Bertini2019, Bertini2019entanglement, Poroli2020dynamics}, and the quantum information locally thrown into the system spreads across the entire system very rapidly.
In the following experiments, we change $W$ and study how much the spread of quantum information is inhibited by increasing the strength of disorder as an indicator of the MBL crossover.

\section{Method}
\label{sec:method}

We utilize the OTOCs and effective quantum volume as indicators of the MBL crossover.
Both quantify the effective size of the causal cone of the quantum circuit. 

\subsection{Out-of-time-ordered correlators}

An OTOC quantifies the operator growth, that is, the increasing of support of a butterfly operator $O_D$ under the unitary evolution  $U$, measuring the noncommutativity with a disjoint operator $O_A$.
The OTOC is defined as 
\begin{equation}
\label{eq:def_OTOC}
\begin{split}    
    \mathrm{OTOC}
    := &
    \Tr\big[\rho U^\dag O_D^\dag U O_A^\dag U^\dag O_D U O_A \big]
    \\
    = &\langle 0^N| 
    (U_\mF)^{\dag n} X_1 (U_\mF)^{n} Z_m (U_\mF)^{\dag n} X_1 (U_\mF)^{n}
    |0^N\rangle ,
\end{split}
\end{equation}
where $\rho$ is a density operator, and $O_A$ and $O_D$ are operators supported on subsystems $A$ and $D$, respectively. In particular, we consider $|0^N\rangle\langle 0^N|$ as a density operator $\rho$, and take $U=(U_\mF)^{n}$, $O_D=X_1$ and $O_A=Z_m$ as shown in the second line of Eq.~(\ref{eq:def_OTOC}). We fix the position of the butterfly operator $O_D$, and move $O_A$ at each time interval $nT$ to measure the operator growth. For this type of OTOCs, where the initial state is an eigenstate of $O_A$~\cite{Liang2024}, we do not need the interferometric circuit to compute OTOCs~\cite{Swingle2016, Swingle2018} that demands a non-local controlled operation for the Hadamard test. Indeed, this may be challenging in NISQ devices supporting only local connectivities between qubits. Instead, we can compute OTOCs by measuring $Z_m$ for the state $|\psi\rangle=(U_\mF)^{\dag n} X_1 (U_\mF)^{n}|0^N\rangle$. Importantly, this can be done in all the sites simultaneously for each time interval $nT$.

To mitigate the noise effect, following 
Refs.~\cite{Swingle2018,Mi2021,Seki2025}, we calculate the normalized OTOC as 
\begin{equation}
\label{eq:normalized_OTOC}
\overline{\mathrm{OTOC}}=    \frac{
        \langle 0^N| (U_\mF)^{\dag n} X_1 (U_\mF)^{n} Z_m (U_\mF)^{\dag n} X_1 (U_\mF)^{n} |0^N\rangle
    }{
        \langle 0^N| (U_\mF)^{\dag n} I_1 (U_\mF)^{n} Z_m (U_\mF)^{\dag n} I_1 (U_\mF)^{n} |0^N\rangle
    }.
\end{equation}
where $I_1$ denotes the identity operator of the qubit that the butterfly operator acts on. 
We note that the denominator takes the value 1 in the absence of errors.
Finally, we take the disorder average of the normalized OTOC that we denote $\langle\langle\overline{\mathrm{OTOC}}\rangle\rangle$.

\subsection{Effective quantum volume}
The denominator of the normalized OTOC~\eqref{eq:normalized_OTOC} can also quantify the growth of the causal cone in the presence of errors. 
To understand the effect of errors, we assume the depolarizing channel model, which was shown to approximate the quantum circuit to compute OTOCs in the quantum chaotic regime very well (See e.g., Ref.~\cite{Seki2025}). 
In the depolarizing channel model, the denominator reads~\cite{Dalzell2024}
\begin{align}
  F &=     \langle 0^N| (U_\mF)^{\dagger n} I_1 (U_\mF)^{n} Z_m (U_\mF)^{\dagger n} I_1 (U_\mF)^{n} |0^N\rangle
  \nonumber \\
\label{eq:quantum_volume}
    &     = (1-p)^{V_{\rm eff}},
\end{align}
where $F$ and $V_{\rm eff}$ are the fidelity and effective quantum volume, respectively~\cite{Kechedzhi2024}.
$V_{\rm eff}$ can be understood as the effective number of two-qubit gates that involve computing the expectation.
$p$ is the error rate of the two-qubit gates (we here assume the error rates are the same in all native two-qubit gates).
The decay of the fidelity becomes large if the operator growth is fast since more two-qubit gates involve computing the expectation value as the size of the light cone increases. 
We consider the disorder average of $F$ (to be precise, the disorder average of the logarithm of $F$) as an indicator of the MBL crossover
since the small $V_{\rm eff}$ implies that the butterfly operator is localized during time evolution. 
We denote $\langle\langle V_{\rm eff}\rangle\rangle=\langle\langle \log F/\log(1-p)\rangle\rangle$, where $\langle\langle \cdot \rangle\rangle$ means the disorder average.

\section{Experiments}
\label{sec:experiments}

We performed experiments to compute OTOCs using IBM's Heron quantum processor named ibm\_fez that consists of 156 superconducting qubits. 
We used $60$ qubits from $156$ qubits that compose the $2\times3$ heavy-hex lattice as shown in Fig.~\ref{fig:qubit_layout}.
The median CZ, SX, and readout errors were typically $3.7\times10^{-3}$, $2.93\times10^{-4}$, and $1.76\times10^{-2}$, respectively, when we submitted the jobs.
We used Qiskit's estimator~\cite{qiskit_paper} to measure $Z_m$. Measurements were performed simultaneously for all sites at each time interval and the realization of the disorder. 
We simulated $25$ realizations of disorder and $10$ Trotter steps, that is, $25\times10=250$ circuits were simulated for each $W$.
The number of shots in each quantum circuit was set to $16,000$. We enabled Pauli twirling for two-qubit gates and for mitigating readout errors named twirled readout error eXtinction (TREX), while dynamical decoupling was disabled. 
The number of twirled circuits (\texttt{num\_randomizations} in Qiskit's twirling options) was set to $32$. 
Unless otherwise stated, we did not use ZNE to compute the denominator and numerator of the normalized OTOCs. 
We discarded the data set for which the denominator of normalized OTOCs in Eq.~\eqref{eq:normalized_OTOC}, that is, the fidelity became negative when computing the disorder average (we discarded unreliable data from the average).
We have checked that this does not change mean values but reduces statistical error bars.

The Trotter circuit denoted by $(U_{\mF})^n$ was implemented by following Ref.~\cite{Kim2023}.
The single $X$ and $Z$ rotation layers were followed by three $R_{ZZ}$ gate layers, and those layers were repeated $n$ times (plus $(U_{\mF})^{\dag n}$ to compute OTOCs).
We fixed the rotation angle of the $R_{ZZ}$ gates by $\frac{\pi}{2}$, and each $R_{ZZ}$ gate was transpiled to one native two-qubit gate.

Furthermore, we optimized the circuits by manually removing unnecessary gates outside of the causal cones as follows. 
We first considered the quantum circuits to compute $\langle 0^N| (U_\mF)^{\dag n} X_1 (U_\mF)^{n} |0^N\rangle$, and manually eliminated redundant gates from $(U_\mF)^{n}$. The resulting optimized operator is denoted as $(U_\mF)^{n}_{\rm opt}$. 
Next, we constructed $(U_\mF)_{\rm opt}^{\dag n}$ from $(U_\mF)_{\rm opt}^{n}$ by utilizing \texttt{inverse} implemented in Qiskit's quantum circuit class. 
Finally, we constructed the target circuits as $(U_\mF)_{\rm opt}^{\dag n} X_1 (U_\mF)_{\rm opt}^{n} |0^N\rangle$ and $(U_\mF)_{\rm opt}^{\dag n} I_1 (U_\mF)_{\rm opt}^{n} |0^N\rangle$ using them. 
In short, the gates inside the causal cone of $X_1$ were taken into consideration. 
This greatly improved the quality of error mitigation, and may be inevitable in practical simulations in present noisy devices.

\begin{figure*}[t]
  \includegraphics[width=.9\textwidth]
  {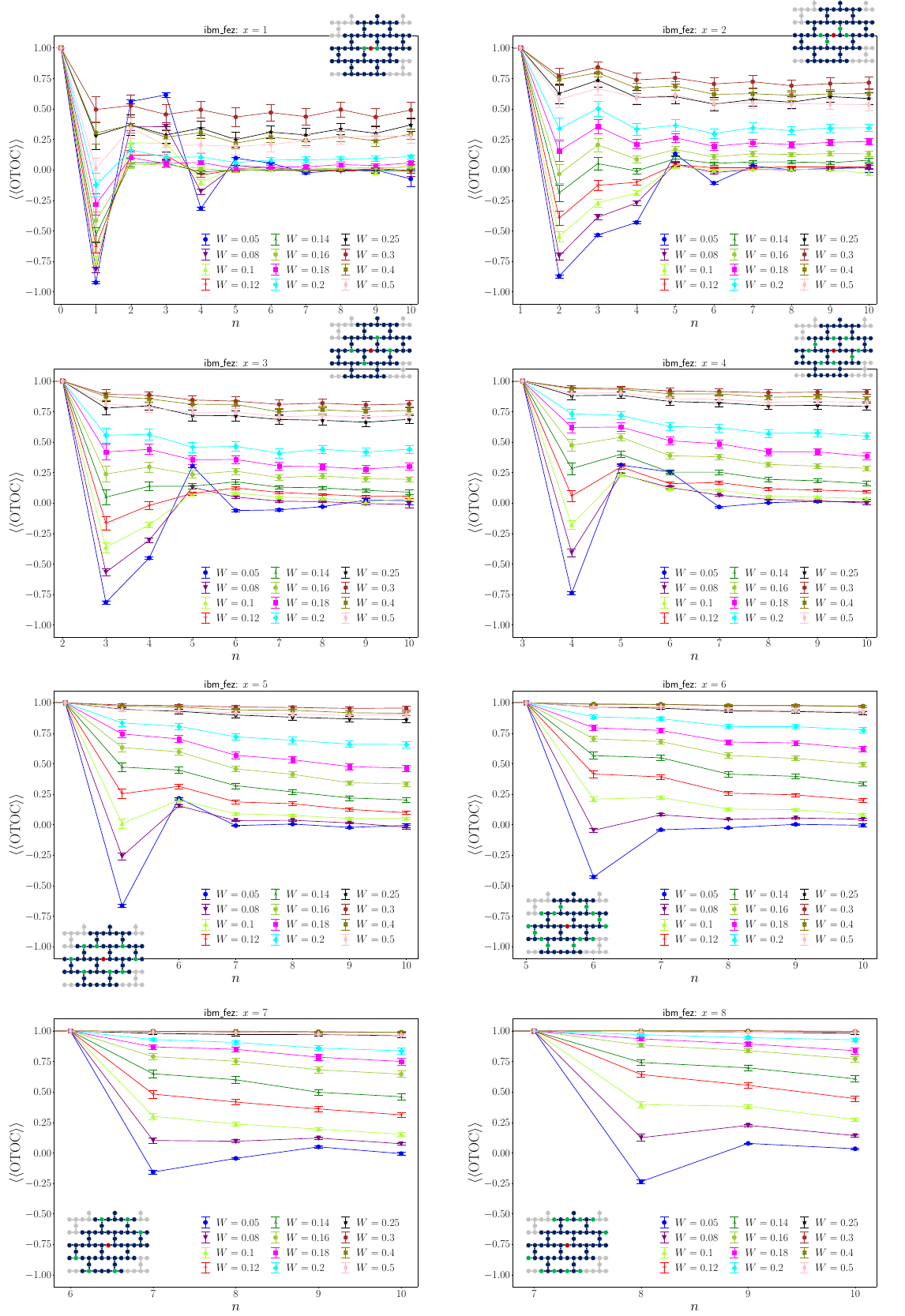}
  \caption{
    \label{fig:OTOC}
    Disorder and site average of the normalized OTOCs defined in Eq.~\eqref{eq:normalized_OTOC} as a function of the number of the Trotter steps. 
    Inset qubit layouts show the positions of $Z_m$, which are colored in green, and $x$ is the distance of the $m$th qubit measured from the qubit that the butterfly operator $X_1$ acts on. 
    The average of the normalized OTOCs over $m$ with the same distance $x$ is taken simultaneously with the disorder average.
    }
\end{figure*}%
\begin{figure}[t]
  \includegraphics[width=.48\textwidth]
  {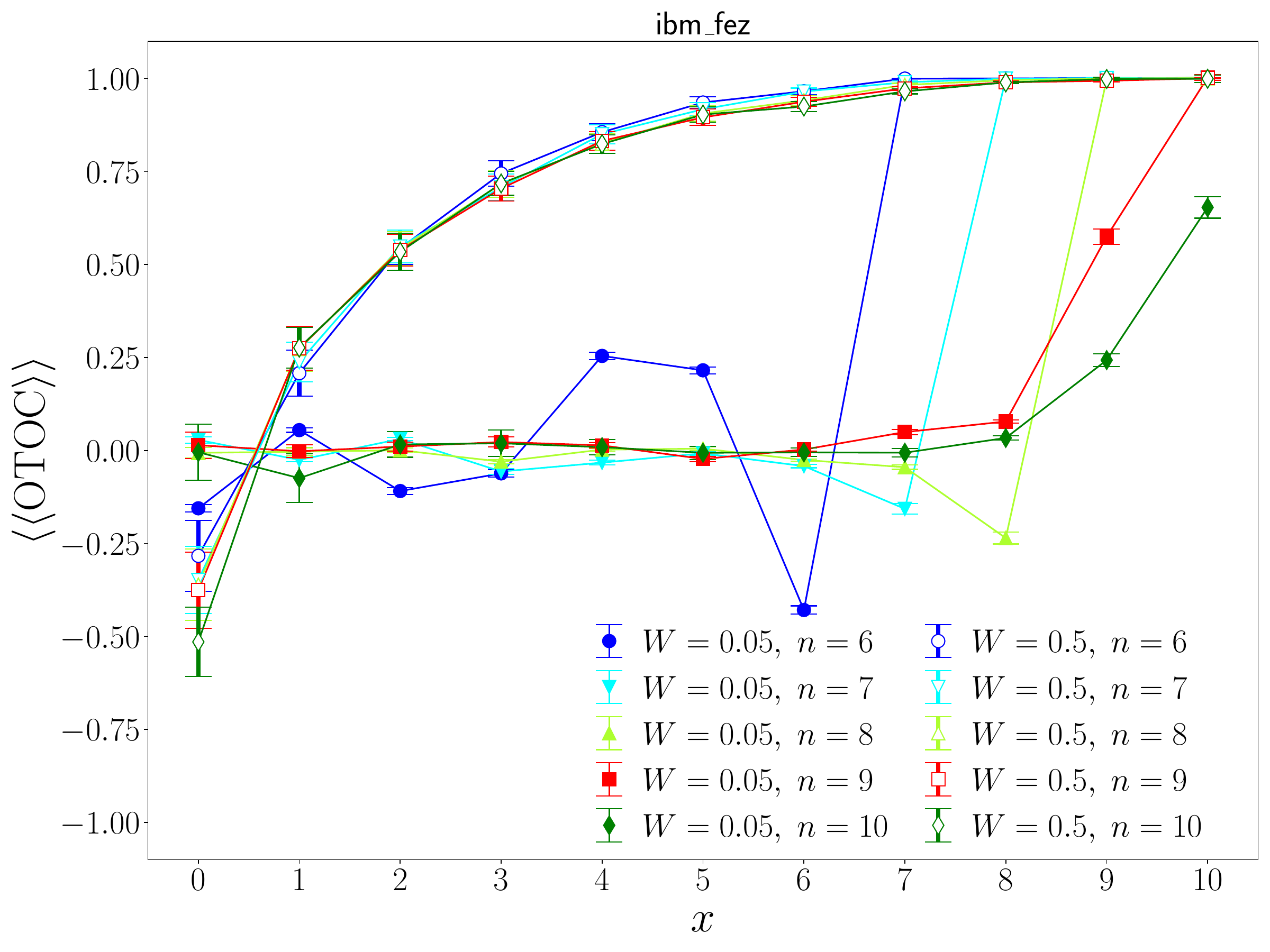}
  \caption{
    \label{fig:OTOC3}
    Disorder and site average of the normalized OTOCs defined in Eq.~\eqref{eq:normalized_OTOC} as functions of the distance of the measured qubits $Z_m$ from the butterfly operator $X_1$ $x$.
    }
\end{figure}%
\begin{figure}[t]
  \includegraphics[width=.48\textwidth]
  {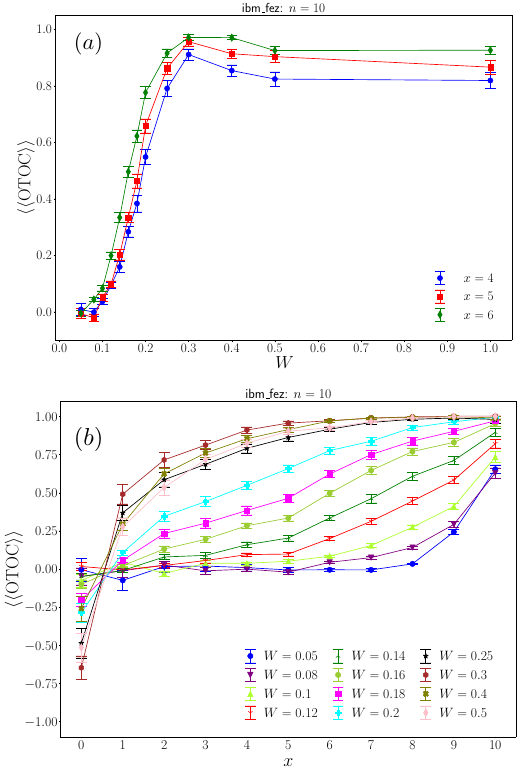}
  \caption{
    \label{fig:OTOC2}
    Disorder and site average of the normalized OTOCs defined in Eq.~\eqref{eq:normalized_OTOC} as functions of (a) the disorder strength $W$ and (b) the distance $x$ of the measured qubits $Z_m$ from the butterfly operator $X_1$.
    }
\end{figure}%

\subsection{Out-of-time-ordered correlators}
\label{sec:otoc}

We show the Trotter step $n$ dependence of the OTOCs for various values of the distance $x$ and the disorder strength $W$ in Fig.~\ref{fig:OTOC}.
Here, the distance $x$ is defined as the shortest path between the butterfly qubit 1 and the measured qubit $m$.  
Since OTOCs depend only on the distance $x$ after the disorder average, we show the disorder and site average of the normalized OTOCs in Fig.~\ref{fig:OTOC}.
The mean and error bars are estimated from the weighted average over the disorder realizations using the measurement errors of $Z_m$ in each disorder realization as weights. 
For small $W$, OTOCs show a sharp drop at $n=x$ and strongly decay at late times.
Those behaviors are consistent with OTOCs in the geometrically local chaotic systems~\cite{Seki2025}. 
For large $W$, on the other hand, they remain around one or show a very slow decay. 
Those behaviors are consistent with the many-body localized phase, and can be interpreted as a slow operator growth of the butterfly operator in the Heisenberg picture, $X_1(nT)=(U_\mF)^{\dag n} X_1 (U_\mF)^{n}$. 
Namely, the support of $X_1(nT)$ does not grow significantly in the localized phase. 
As a result, $X_1(nT)$ almost commutes with $Z_m$ if $m\neq1$, and OTOCs remain close to one. 

In Fig.~\ref{fig:OTOC3}, we show the late-time ($n\geq 6$) behaviors of the OTOCs as a function of $x$ at $W=0.05$ and $W=0.5$ as representatives of small and large disorder strengths, respectively.
On one hand, we see a clear wave front of OTOCs around $x=n$ for $W=0.05$.
On the other hand, OTOCs are almost insensitive to Trotter steps $n$ for $W=0.5$, implying that information does not spread significantly when the disorder is strong.

To locate the ergodic and the MBL regimes as a function of the disorder strength $W$, we now estimate a characteristic disorder strength $W_c$, hereafter referred to as a crossover point.
To this end, as an indicator of the MBL crossover, we select the OTOCs with the largest Trotter step $n=10$ studied at the halfway point $x=\frac{n}{2}=5$ due to the following reasons.
(i) Our available Trotter step $n$ is limited due to unavoidable noises in present devices, and we cannot see the asymptotic behaviors of OTOCs for large $x$, e.g., for $x=7,8$ as seen in Fig.~\ref{fig:OTOC}.
(ii) OTOCs with small $x$, on the other hand, may not reflect the many-body effects in the large system size correctly due to their shallow causal cones (the system size may be effectively small).
The expected behaviors of the OTOCs with $x=\frac{n}{2}$ in the two extreme cases are as follows.
In the extremely chaotic regime, the OTOCs with $x=\frac{n}{2}$ are expected to be almost zero since the distance $x=\frac{n}{2}$ is deep inside the causal cone.
In the extremely many-body localized regime, the OTOCs with $x=\frac{n}{2}$ are expected to be almost one since the growth of the butterfly operator is so slow that the qubits at distance $x=\frac{n}{2}$ remain unaffected by the information thrown at the qubit 1. 

Figure~\ref{fig:OTOC2}(a) shows the disorder strength $W$ dependence of the renormalized OTOCs at $x=\frac{n}{2}=5$.
The experimental data clearly show a crossover from the chaotic regime (strong decay of OTOCs) to the many-body localized regime (weak decay of OTOCs) as a function of $W$,
where the height of the OTOCs behaves like an order parameter of spontaneous symmetry breaking.
We also show the OTOCs with $x=\frac{n}{2}\pm1=4,6$ for comparison in Fig.~\ref{fig:OTOC2}(a).
We find that the qualitative behavior of the OTOCs is the same for those points.
We now define the crossover point $W_c$ as the disorder strength at which the OTOCs take the largest slope, that is, $W_{c} = \underset{W}{\arg \max}\ \partial\langle\langle\mathrm{OTOC}\rangle\rangle/\partial W$.
Accordingly, we locate the ergodic and the MBL regimes below and above $W_c\sim0.18$, respectively.  

We also try to estimate $W_c$ from another information probed by OTOCs to deepen our understanding in the behavior of OTOCs around $W_c$.
We show the distance $x$ dependence of the OTOCs at $n=10$ in Fig.~\ref{fig:OTOC2}(b).
We see that the behavior of the wavefront of OTOCs changes qualitatively as $W$ increases.
We find a sharp wave front at $x=n$ for small $W$, 
while it becomes broader as $W$ increases and finally we cannot find the wave front (See also Fig.~\ref{fig:OTOC3} for the behavior of the wavefront in the weak and strong disorder regimes).
Those qualitative behaviors of the wavefront change around $W=0.18$.
This estimation of the crossover point is consistent with the one from the slope of OTOCs with $x=\frac{n}{2}$ as a function of $W$.

Rigorously speaking, we need the finite-size scaling analysis to discuss whether there is a genuine phase transition between the ergodic and MBL phases or it remains a crossover.
Although we may have a sufficient number of qubits, we cannot run the circuits with such a large Trotter step as the causal cone of the butterfly operator reaches all qubits. 
This can be understood by estimating the decay of fidelity.
We leave the finite-size scaling analysis for future work.

\begin{figure*}[t]
  \includegraphics[width=.9\textwidth]
  {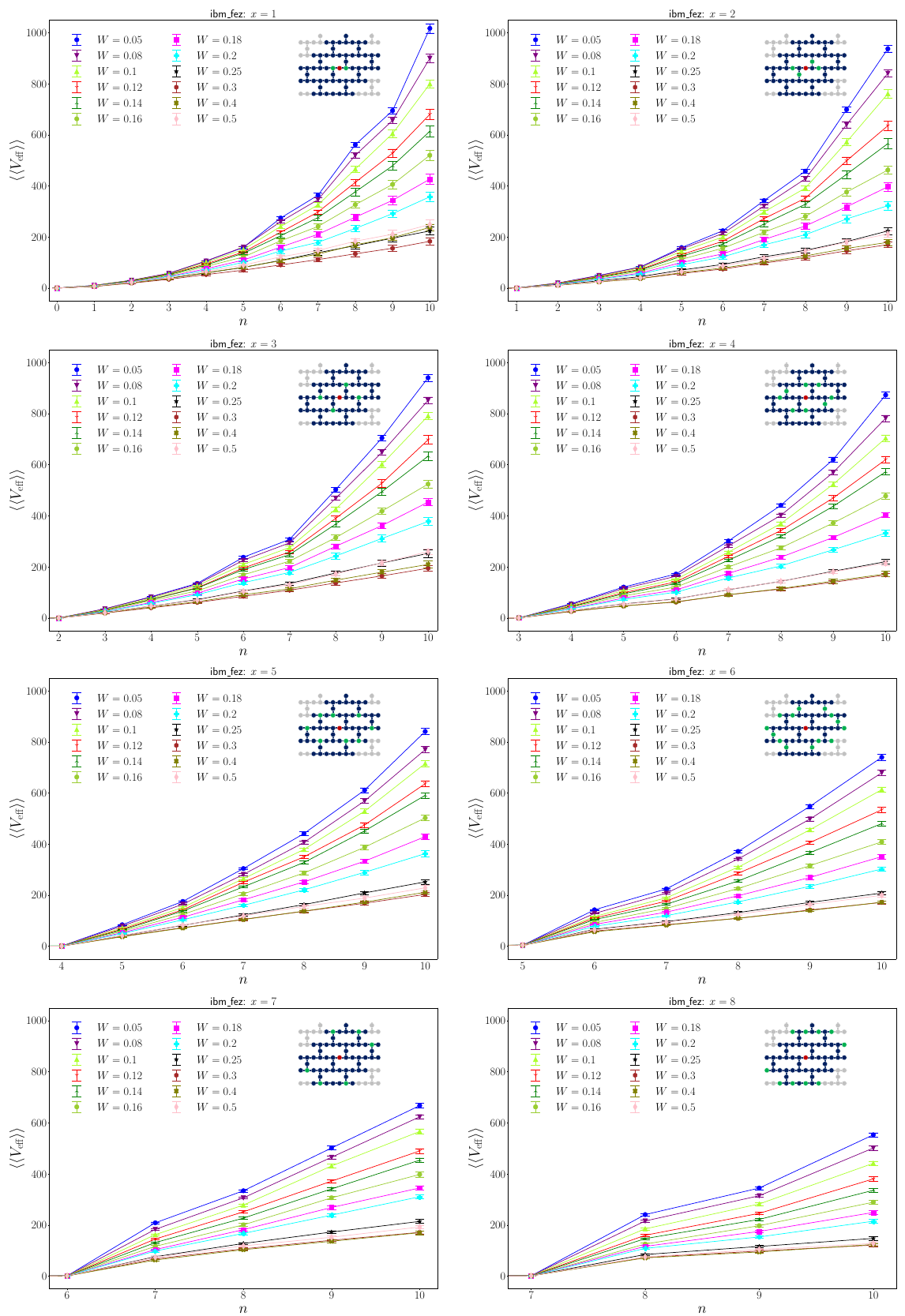}
  \caption{
    \label{fig:V}
    Disorder and site average of the effective quantum volume defined in Eq.~\eqref{eq:quantum_volume} as a function of the number of the Trotter steps. 
    Inset qubit layouts show the positions of $Z_m$, which are colored in green, and $x$ is the distance of the $m$th qubit measured from the qubit that the butterfly operator $X_1$ acts on. 
    The average of the effective quantum volume over $m$ with the same distance $x$ is taken simultaneously with the disorder average.
    }
\end{figure*}%
\begin{figure}[t]
  \includegraphics[width=.48\textwidth]{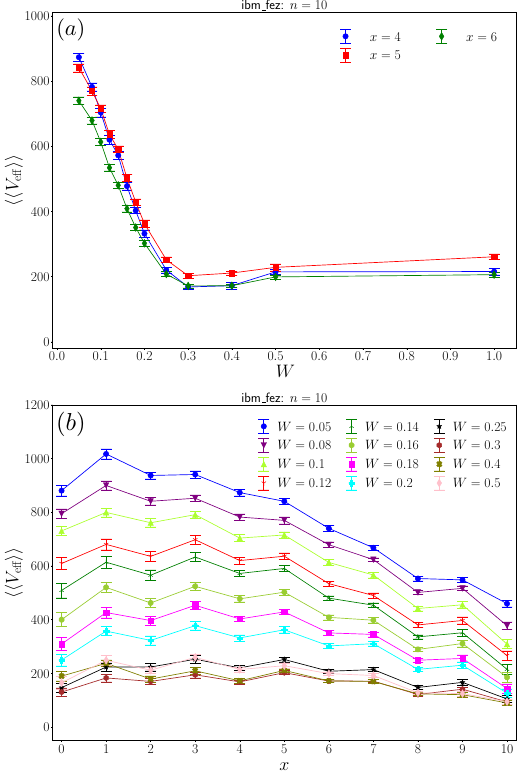}
  \caption{
    \label{fig:V2}
    Disorder and site average of the effective quantum volume defined in Eq.~\eqref{eq:quantum_volume} as functions of (a) the disorder strength $W$ and (b) the distance $x$ of the measured qubits $Z_m$ from the butterfly operator $X_1$.
    }
\end{figure}%

\subsection{Effective quantum volume}
\label{sec:denominator}
The denominator of the normalized OTOCs in Eq.~\eqref{eq:normalized_OTOC}, $\langle 0^N| (U_\mF)^{\dag n} I_1 (U_\mF)^{n} Z_m (U_\mF)^{\dag n} I_1 (U_\mF)^{n} |0^N\rangle$ 
can be useful not only as a calibrator in error mitigation, but also as a probe of the growth of the causal cone.
We show the Trotter step $n$ dependence of the effective quantum volume $V_{\rm eff}$ in Fig.~\ref{fig:V}.
We again consider the site average of the effective quantum volume as well as the disorder average.
The mean and error bars are estimated from the weighted average over the disorder realizations using the measurement errors of $Z_m$ in each disorder realization as weights. 
We see that the effective quantum volume grows superlinearly for small $W$ at late times, 
while it grows linearly for large $W$ at late times.
Due to a shortage of the number of Trotter steps, we cannot perform the quantitative analysis in the chaotic phase such as the estimation of the asymptotic scaling of the effective quantum volume $V_{\rm eff}\sim n^a$.
On the other hand, the linear growth at late times implies that the causal cone of the butterfly operator does not grow 
(The constant causal cone results in the linear growth of the effective quantum volume as a function of the Trotter steps).
This gives a clear indication of the many-body localized phase.

As seen above, the effective quantum volume is useful to probe the MBL regime.
However, it is not useful for the quantitative analysis of the crossover point.
We show the disorder strength $W$ dependence of the effective quantum volume at late time $n=10$ in Fig.~\ref{fig:V2}(a).
We also show the distance $x$ dependence of the effective quantum volume at $n=10$ in Fig.~\ref{fig:V2}(b).
We find that there are two phases in the effective quantum volume as a function of $W$.
However, we do not find a rapid change of the effective quantum volume, that is, the crossover point in Fig.~\ref{fig:V2}(a) nor the wave front in Fig.~\ref{fig:V2}(b).

\subsection{Quality of error mitigation}
\label{sec:denominator}
In previous studies~\cite{Mi2021,Seki2025}, it was shown that the error mitigation using renormalization is effective in the chaotic phase.
We here check the validity of the error mitigation outside of the chaotic phase.
To this end, we run the circuits to compute the renormalized OTOCs with amplifying noises.
We use the gate folding techniques implemented in Qiskit's ZNE~\cite{qiskit_paper} to amplify the two-qubit gate errors, which would be the major sources of errors in the present devices.  
We set \texttt{noise\_factors} in Qiskit's ZNE options to $1.0$ and $1.5$, which is the noise amplification factor, and we denote $f$ in the following.
The noise factor $f$ means that the noise is amplified by $f$ times, that is, the two-qubit gate errors are on average set to $fp$ with $p$ being the original error rate.
We compute the renormalized OTOCs defined in Eq.~\eqref{eq:normalized_OTOC} with the two noise factors $f=1.0$ and $f=1.5$.
If the error mitigation by renormalization is effective, the renormalized OTOCs should be resilient against the change of $f$.
By checking this, we can verify the validity of the error mitigation only by using experimental data.

\begin{figure}[t]
  \includegraphics[width=.48\textwidth]
  {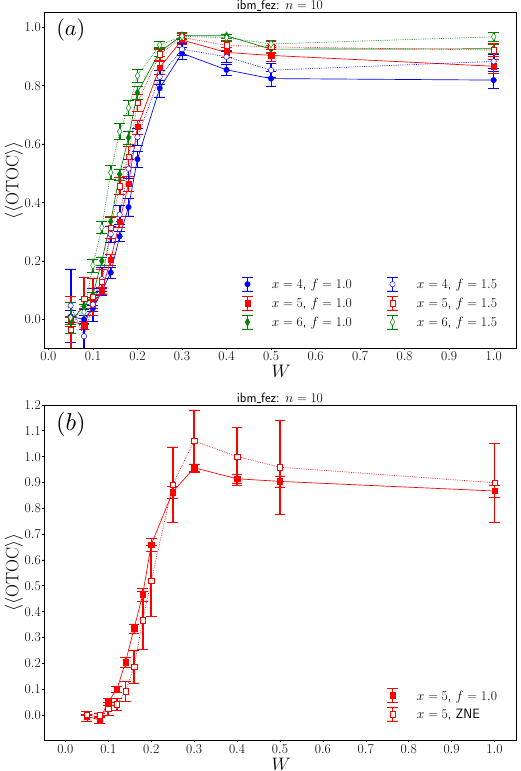}
  \caption{
    \label{fig:ZNE}
    (a) Disorder and site average of the normalized OTOCs defined in Eq.~\eqref{eq:normalized_OTOC} as a function of the disorder strength $W$ with the noise factor $f=1.0$ and $1.5$. 
    (b) Disorder and site average of the normalized OTOCs defined in Eq.~\eqref{eq:normalized_OTOC} with the noise factor $f=1.0$ and OTOCs estimated using ZNE of the bare OTOCs defined in Eq.~\eqref{eq:def_OTOC}. 
    The experimental data of the noise factor $f=1.0$ and $1.5$ are used in ZNE. 
  }
\end{figure}

We show the disorder strength $W$ dependence of the renormalized OTOCs at $x=4,5,6$, and $n=10$ in Fig.~\ref{fig:ZNE}(a).
We find that the qualitative behavior is the same in $f=1.0$ and $1.5$, 
but the height of the renormalized OTOCs seems to be quantitatively larger for $f=1.5$ than that for $f=1.0$.
The difference is within the error bar, and we may rely on error mitigation by the renormalization in the chaotic and many-body localized phases.
However, it may be beyond the error bar around the crossover point ($0.14\leq W \leq 0.2$), and we may need the careful analysis in this region.
To this end, we perform the ZNE using the bare OTOCs defined in Eq.~\eqref{eq:def_OTOC}.
We fit the bare OTOCs at $f=1.0$ and $1.5$ by exponential functions of the form $a {\rm e}^{-bf}$, where $a$ is the estimation of OTOCs from ZNE.
We note that we do not use the information of the denominator of the normalized OTOCs in ZNE, so that we can use those methods as a cross-check.

\begin{figure}[t]
  \includegraphics[width=.48\textwidth]
  {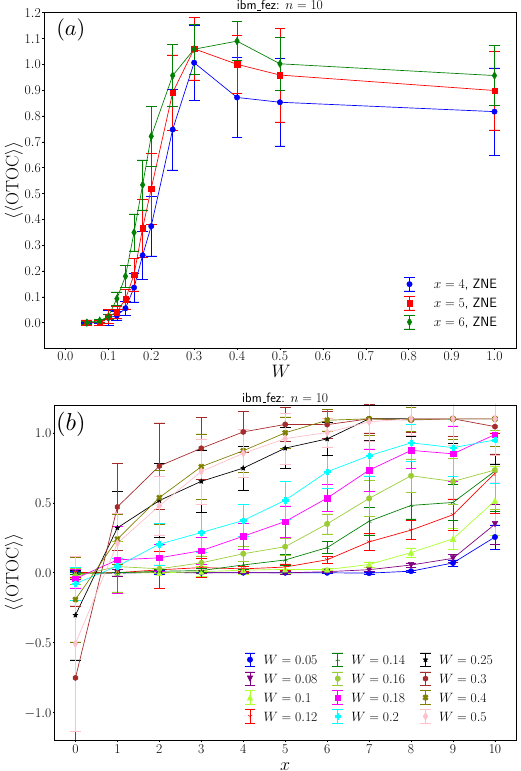}
  \caption{
    \label{fig:ZNE2}
    Disorder and site average of the OTOCs estimated using ZNE as functions of (a) the disorder strength $W$ and (b) the distance $x$ of the measured qubits $Z_m$ from the butterfly operator $X_1$.
    The experimental data of the bare OTOCs defined in Eq.~\eqref{eq:def_OTOC} with the noise factor $f=1.0$ and $1.5$ are used in ZNE. 
  }
\end{figure}

We show the renormalized OTOCs and OTOCs estimated using ZNE at $x=5$ and $n=10$ in Fig.~\ref{fig:ZNE}(b).
As expected, those estimated OTOCs are consistent within the error bars in the chaotic and the MBL regimes. 
The slope of OTOCs around the crossover point becomes steeper with ZNE, and the phase-transition like behavior gets clearer.
Finally, we do the same analysis shown in Fig.~\ref{fig:OTOC2} using ZNE.
We show the disorder strength $W$ dependence of the OTOCs estimated by ZNE at $n=10$ in Fig.~\ref{fig:ZNE2}(a).
We also show the distance $x$ dependence of the OTOCs at $n=10$ in Fig.~\ref{fig:ZNE2}(b).
We find that the qualitative behavior of OTOCs is the same for those estimations although the error bars in ZNE are relatively large, and the estimation by the renormalization may be more reliable in small or large $W$.
We again observe a crossover from the chaotic phase (strong decay of OTOCs) to the many-body localized phase (weak decay of OTOCs) in Fig.~\ref{fig:ZNE2}(a).
We may estimate the MBL crossover point from the slope of the OTOCs as $W_c\sim0.18$ and consistent with the estimation from the previous analysis.


\section{Discussion}
\label{sec:conclusion}
We have simulated the MBL crossover of the disordered kicked-Ising model on a heavy-hex lattice using $60$ qubits from $156$ qubits available in the IBM Heron r2 device named ibm\_fez. 
To this end, we computed OTOCs as an indicator of the MBL crossover. 
We used the late-time behavior of OTOCs to characterize the quantum chaotic and the MBL regimes, and located the two regimes as a function of the disorder strength. 
The validity of the results was confirmed by comparing two error mitigation methods, that is, the operator renormalization method and zero-noise extrapolation.

While we have successfully observed qualitative differences in the spread of the disorder-averaged OTOC between strong- and weak-disorder cases up to 10 Trotter steps, we did not conduct the longer-time simulations due to the limitations of noise. 
Theoretically and numerically, it is shown for 1D MBL systems that the OTOCs exhibit logarithmic light cones whose spatial extent increases only logarithmically in time~\cite{Huang2016,Chen2016,ChenXiao2016,Fan2017}. 
Experimentally exploring the extent of light cones in 2D or higher dimensional systems can be a future direction using digital quantum computers with lower noise levels.

We have not discussed the property of the MBL transition because we have conducted the simulation only in one system size.
However, it should be noted that a stable MBL phase in the thermodynamic limit has been challenged in recent years~\cite{Sierant_2025,li2025,hur2025}.
If the MBL phase is unstable in the thermodynamic limit, there should be only a  crossover instead of a transition, and what we observed in the experiments is a prethermal phenomenon.
We should also note that the weaker finite-size drift of $W_c$ has been observed in one-dimensional kicked-Ising models, suggesting still the possibility of a stable MBL phase in Floquet systems~\cite{Sierant2023}. 
It is considered that the lack of energy conservation plays an important role in stabilizing the MBL phase.
Furthermore, the stability of the MBL phase remains an open question in two and higher spatial dimensions.
Considering these facts, further studies of the MBL transition in two- or higher-dimensional Floquet many-body systems are very important, but may be hard using classical computers.
Digital quantum computers will be new useful platforms for them as well as studies of other nonequilibrium phases of quantum many-body systems.

 \section*{Data Availability}
 The experimental data that support the findings of this study are available at Zenodo~\cite{Zenodo}.

 \section*{Code Availability}
 The code that supports the findings of this study is available from the corresponding author upon reasonable request.
\\

\begin{acknowledgments}
T. H. thanks Yuta Kikuchi for fruitful discussions and critical comments.
A portion of this work is based on results obtained from project JPNP20017, subsidized by the New Energy and Industrial Technology Development Organization (NEDO). 
T.~H. is also supported by JSPS KAKENHI Grants No. JP24K00630 and No. JP25K01002.
K.~S. is also supported by JSPS KAKENHI Grants No. JP22K03520.
S.~Y. is also supported by JSPS KAKENHI Grants No. 21H04446 and No. JP24K02948, MEXT for the Program for Promoting Research of the Supercomputer Fugaku (Grant No. MXP1020230411), 
and the Center of Excellence (COE) Research Grant in Computational Science from Hyogo Prefecture and Kobe City through the Foundation for Computational Science.
We are also grateful for the funding received from JST COI-NEXT (Grant No. JPMJPF2221).  
Additionally, we acknowledge the support from the UTokyo Quantum Initiative, and the RIKEN TRIP initiative (RIKEN Quantum).
\end{acknowledgments}

\bibliography{scrambling}
\end{document}